\documentclass[preprint2]{aastex}

\shorttitle{The dayside spectrum of 51 Peg b}
\shortauthors{Brogi et al.}

\begin{document}

\title{Detection of molecular absorption \\ in the dayside of exoplanet 51 Pegasi b?}


\author{M. Brogi\altaffilmark{1}, I. A. G. Snellen\altaffilmark{1}, R. J. de Kok\altaffilmark{2}, S. Albrecht\altaffilmark{3}, J. L. Birkby\altaffilmark{1} \and E. J. W. de Mooij\altaffilmark{4}}
\email{brogi@strw.leidenuniv.nl}


\altaffiltext{1}{Leiden Observatory, Leiden University, P.O. Box 9513, 2300 RA Leiden, The Netherlands}
\altaffiltext{2}{SRON Netherlands Institute for Space Research, Sorbonnelaan 2, 3584 CA Utrecht, The Netherlands}
\altaffiltext{3}{Department of Physics, and Kavli Institute for Astrophysics and Space Research, Massachusetts Institute of Technology, Cambridge, Massachusetts 02139, USA}
\altaffiltext{4}{Department of Astronomy and Astrophysics, University of Toronto, 50 St George Street, Toronto, Ontario M5S 3H4, Canada}


\begin{abstract}
In this paper we present ground-based high-resolution spectroscopy of 51 Pegasi using CRIRES at the Very Large Telescope. The system was observed for 3$\times$5 hours at 2.3 $\micron$ at a spectral resolution of R = 100,000, targeting potential signatures from carbon monoxide, water vapour and methane in the planet's dayside spectrum. In the first 2$\times$5 hours of data, we find a combined signal from carbon monoxide and water in absorption at a formal 5.9$\sigma$ confidence level, indicating a non-inverted atmosphere. We derive a planet mass of $M_\mathrm{P} = (0.46 \pm 0.02) M_\mathrm{Jup}$ and an orbital inclination $i$ between $79.6^\circ$ and $82.2^\circ$, with the upper limit set by the non-detection of the planet transit in previous photometric monitoring. However, there is no trace of the signal in the final 5 hours of data. A statistical analysis indicates that the signal from the first two nights is robust, but we find no compelling explanation for its absence in the final night. The latter suffers from stronger noise residuals and greater instrumental instability than the first two nights, but these cannot fully account for the missing signal. It is possible that the integrated dayside emission from 51 Peg b is instead strongly affected by weather. However, more data are required before we can claim any time variability in the planet's atmosphere.
\end{abstract}


\section{Introduction}

The first discovery of an exoplanet around a main-sequence star, the G-dwarf 51 Pegasi, was announced in 1995 (\citeauthor{may95}). The presence of the planet was inferred from the measurement of a periodic Doppler-shift in the stellar lines, which was interpreted as being due to the motion of the star around the center of mass of the star-planet system. Since then, the radial velocity technique has been revolutionary in exoplanet research, resulting in hundreds of discoveries.

51 Pegasi b is unlike any planet in our own Solar System: with a minimum mass of $\sim$~0.5 $M_\mathrm{Jup}$, it orbits at only $\sim$~0.05 AU from the star, well within the orbit of Mercury (0.47 AU). At the time of discovery, the generally accepted theories of planet formation and evolution could not explain the presence and survival of a Jupiter-size planet at such short orbital distance. This resulted in an intense debate on whether the radial velocity signal of 51 Peg really was caused by an unseen planet.

The first concern was that, since the orbital inclination $i$ was unknown and therefore the measured mass was only a lower limit, 51 Pegasi b could actually be a low-mass star or brown dwarf seen face-on. This hypothesis was rejected because of the low probability of this geometrical configuration (2.5$\times 10^{-5}$ for the hydrogen-burning limit of 0.08 $M_\odot$), the stellar projected rotational velocity \citep{fra96}, the absence of an X-ray signal from the system \citep{pravdo96}, and eventually the discovery of other 51 Pegasi-like objects \citep{mar96,but97}. At the same time, \citet{lin96} showed that 51 Pegasi b could have formed much further away from the parent star, and subsequently migrated inward through tidal interaction with the protoplanetary disk. Moreover, \citet{rasio96} showed that the timescale for orbital decay via tidal dissipation was possibly longer than the main-sequence lifetime of the star.

Another concern was that stellar radial pulsations or hot spots in the surface could mimic the Doppler signal induced by a short-period giant planet. This hypothesis was also found to be unlikely, because the slow stellar rotation and low chromospheric activity are inconsistent with a high-amplitude signal with a period of a few days. Nevertheless, \citet{gray97} performed a bisector analysis on the spectral lines of 51 Peg, concluding that the RV signal was possibly caused by a non-radial, unknown mode of stellar oscillation, rather than by a planet. However, subsequent bisector analyses at higher spectral resolution showed no variations \citep{hat97,hat98}. In the meantime, adding confidence to the planet hypothesis, the first exoplanets in eccentric orbits were found, making it difficult to explain the RV signals with stellar mechanisms \citep{mar96}. Eventually, transits of the hot-Jupiter HD 209458 b were observed \citep{hen00,charb00}, finally settling the debate on the planetary nature of radial velocity measurements and opening the era of exoplanet characterization.

Although RVs are very successful at detecting planets, they offer little information about the planet itself, whereas transits reveal the planet radius and the system inclination, which, when combined with the radial velocity curve, determine the planet mass, its mean density, and a constraint on the planet's internal structure. Furthermore, starlight can filter through the planet atmosphere during transit, showing an imprint of its atomic and/or molecular gases. Atmospheric absorption increases the planet effective radius and the transit depth at a given wavelength. Therefore, by measuring the planet radius as a function of wavelength a {\it transmission spectrum} can be constructed. This technique has led to the identification of atomic sodium \citep{charb02, sne08, red08}, potassium \citep{sing11}, hydrogen \citep{vid03}, carbon and oxygen \citep{vid04}, methane and water \citep{tin07,swa08,des09,sing09,gib11}. In addition, the planet is occulted by the star around superior conjunction. Its light is temporarily blocked, allowing its flux to be measured directly by comparison with the out-of-eclipse total system flux. Depending on wavelength, this reveals the planet's thermal emission, possibly modulated by molecular features \citep{gri08,swa09}, and/or reflected starlight \citep{maz12, mis12}. Finally, continuous flux monitoring of a transiting system can reveal the phase function of the planet \citep{knu09,cro10,knu12}.

Until recently, exoplanet atmospheric characterization was limited to transiting planets, leaving out most of the planets discovered via the RV technique. For the first time, \citet{sne10} demonstrated that ground-based high-resolution spectroscopy and RVs can also be used for exoplanet characterization. They detected the absorption signature of carbon monoxide in the transmission spectrum of HD 209458 b, using the CRIRES spectrograph on the ESO Very Large Telescope. Its spectral resolution of R$\sim$100,000 allows the planet signature, which changes in Doppler shift during the observations due to the orbital motion of the planet, to be separated from the stationary telluric and stellar absorption lines. In addition to transmission spectroscopy, this method can be applied to dayside spectroscopy, probing the planet when it is almost fully illuminated, just before and after secondary eclipse. In this case the planet thermal emission, modulated by the atmospheric molecular signature, is directly detected and therefore does not require the planet to transit. This has led to the detection of carbon monoxide absorption in the dayside spectrum of the non-transiting hot Jupiter $\tau$ Bo\"otis b, also revealing its orbital inclination and planet mass \citep{bro12,rod12}. 

In this paper we apply the same method to the non-transiting planet 51 Pegasi b. The reminder of the text is organized as follows: in Section~\ref{obs} we describe our observations and data reduction techniques. The resulting signal is described in Section~\ref{results}, while Section~\ref{discussion} extensively discusses the non-detection in one of the three nights of observations. Assuming that the signal in the first two nights is genuine, Section~\ref{parameters} presents the derived parameters of 51 Pegasi b and of its atmosphere. Finally, a recap of our main findings and the future perspectives of high-resolution ground-based spectroscopy are presented in Section~\ref{concl}.

\section{Observations and Data Reduction}\label{obs}

\subsection{Telescope and instruments}

High-dispersion infrared spectra of 51 Pegasi (K~=~3.91 mag) were taken with the ESO Very Large Telescope (VLT) as part of the Large Program 186.C-0289, aimed at the atmospheric study of the brightest transiting and non-transiting hot Jupiters visible from Cerro Paranal. The system was observed in 2010 on the nights of October 16, 17 and 25 at low airmass, using the CRyogenic Infra-Red Echelle Spectrograph \citep[CRIRES,][]{kau04}, mounted at the Nysmith-A focus of the VLT Antu. The spectra are imaged on four 1024$\times$512 pixels Aladdin II detectors, separated by small gaps of about 100 pixels. At a resolution of R~$\sim$~100,000, 904 spectra were collected during a total of $\sim$~16 hours of observations in the wavelength range 2.287--2.345 $\micron$, which covers the 2--0 R-branch of carbon monoxide. The Multi-Application Curvature Adaptive Optic system \citep[MACAO,][]{ars03} was employed to maximize the throughput of the 0.2$\arcsec$ slit. During each night, the target was observed without interruption while nodding along the slit by 10$\arcsec$ between consecutive spectra, according to an ABBA pattern, to allow a proper background subtraction. A standard set of calibration frames was taken in the morning after each night of observation. 

\subsection{Basic data analysis}\label{datared}

The initial data reduction was performed using the CRIRES pipeline 1.11.0\footnote{The pipeline documentation is available at \url{ftp://ftp.eso.org/pub/dfs/pipelines/crire/}}. Each set of AB or BA spectra was flat-fielded, corrected for bad-pixels and non-linearity and combined in order to subtract the background. One-dimensional spectra were subsequently extracted with the optimal extraction algorithm \citep{hor86}, resulting in 452 spectra. The subsequent data analysis was performed with purpose-built IDL scripts. For each night and each detector, the spectra were handled as a matrix with wavelength (or pixel number) on the $x$-axis and time (or phase) on the $y$-axis, and treated separately, following the procedure in \citet{sne10} and \citet{bro12}. This is necessary because each night of observations has different atmospheric and instrumental conditions, and because each detector is read out using a different amplifier, with particular characteristics that need to be modeled independently. 

In each spectrum, bad pixels and bad regions were corrected through spline and linear interpolation respectively. The identification of bad pixel/regions was first done by eye on each matrix of data. Residual outliers were removed iteratively during the whole reduction sequence. Subsequently, each spectrum was aligned to a common wavelength scale. The difference between the centroids of the telluric lines and their average value across the entire series was fitted with a linear function in pixel position, and the resulting shift was applied via spline-interpolation. After the alignment the typical scatter in the residual position of the line centroids is on average less than 0.1 pixels. The common wavelength solution was determined by fitting a second-order polynomial to the pixel positions of the lines in the average spectrum, as function of their corresponding wavelengths listed in the HITRAN database \citep{rot09}. Finally, each spectrum of the series was normalized by its median value.

\subsection{Stellar subtraction}\label{nostar}

Compared to our previous analysis of HD 209458 and $\tau$ Bo\"otis, an additional step -- the removal of the lines in the stellar spectrum -- was required. This is because 51 Pegasi is a comparatively cooler star with an effective temperature of $T_\mathrm{eff} = (5790 \pm 44)$ K \citep{fis05}, and thus several spectral lines, in particular those of CO, are also present in the stellar spectra. Due to the changing barycentric velocity of the observatory during the night ($\Delta v_\mathrm{obs} \sim$ 0.5 -- 1 km s$^{-1}$), mainly caused by the rotation of the Earth, the stellar lines shift by 0.3 -- 0.6 pixels in wavelength during our observations. If the stellar spectrum is not subtracted from the data, the procedure used to remove the telluric contamination (see Section~\ref{telluric}) would produce strong residuals at the position of the stellar lines. Since the effective temperature of 51 Pegasi is very similar to that of the Sun, a stellar template was created from a high-resolution solar spectrum \citep{atl96}. The position, amplitude and width of the solar lines were fitted with Gaussian or Lorentzian profiles (where appropriate) and a model spectrum was built from the list of the fitted lines and convolved to the CRIRES resolution. Each frame was then divided through the normalized template spectrum, shifted in wavelength according to the radial velocity of the star 51 Pegasi, and multiplied by a factor of 0.95 in order to match the line depths relative to those of the Sun. The need for a scaling factor is possibly due to differences in spectral type, metallicity and/or rotational velocity between the two stars.

\subsection{Removal of the telluric contamination}\label{telluric}

The most dominant feature in the spectra is telluric absorption, which has a fixed pattern in wavelength but is variable in depth due to changes in airmass, atmospheric conditions and instrumental response. In order to minimize the telluric signature, the logarithm of the flux in each column of the spectral series was first linearly fitted with geometric airmass and divided out. This greatly reduced the amplitude of the telluric features, although residuals at the 1--5\% level were still present at the positions of some of the strongest absorption lines, due to second-order effects. For example, possible changes in the water vapour content during the night will generally not follow airmass. The behaviour of the residuals as function of time was therefore measured in a few strong lines and the flux in each column of the matrix was fitted by a linear combination of those measurements, computed via linear regression. In addition, residual large-scale gradients in the matrix were removed by applying a high-pass filter to each spectrum of the series. Finally, each column was divided by the square of its standard deviation, in order to scale down the noisy parts of the spectra according to their signal-to-noise. These would otherwise dominate the cross-correlation analysis performed later.

\begin{figure*}[ht]
\includegraphics[width=16cm]{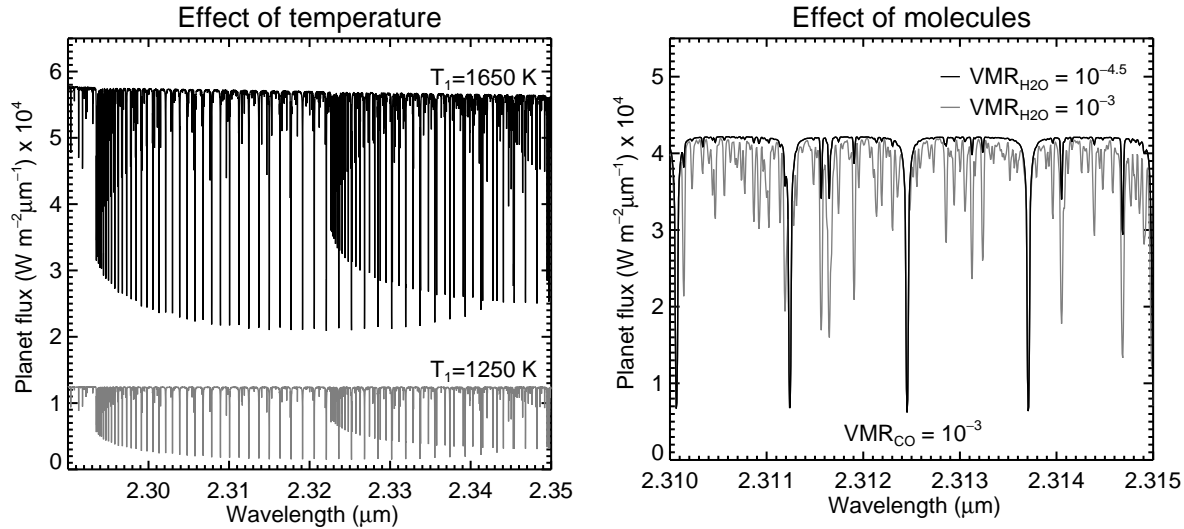}
\caption{Example model spectra for the atmosphere of the planet 51 Pegasi b, showing the effects of different base temperatures $T_1$ (left panel) and different Volume Mixing Ratios of water and carbon monoxide (right panel). Details about these models are given in Section~\ref{atmmodel}.}\label{models}
\end{figure*}

\subsection{Atmospheric models}\label{atmmodel}

In order to be able to combine the spectral lines of 51 Pegasi b via a cross-correlation technique, model spectra of the planet atmosphere were computed assuming an average dayside vertical profile in hydrostatic equilibrium. The atmosphere is assumed to be isothermal for pressures $p > p_1$ and $p < p_2$, at temperatures $T_1$ and $T_2$ respectively. In between the two pressure levels, a constant lapse rate $[d\log(p)]/dT$ is assumed. Inverted and non-inverted $T/p$ profiles are modelled by assuming $T_2 > T_1$ and $T_2 < T_1$ respectively. Such a vertical structure is the simplest possible for a planet atmosphere, and although it is probably not representative for the entire vertical extent, it is a reasonable assumption in the limited range of temperature and pressure probed by our observations. More importantly, given the degeneracies between vertical profiles and molecular abundances (see Section~\ref{atm}), a more complex description of the planet atmosphere would be beyond the purposes of this study. Molecular species were assumed to be vertically mixed, with the dominant sources of opacities at 2.3 micron being carbon monoxide, water and possibly methane. Opacities for CO and H$_2$O were taken from the HITEMP database \citep{rot10}, while those for CH$_4$ were taken from HITRAN \citep{rot09}. A grid of models was computed assuming volume mixing ratios (VMRs) between $-4.5 < \log_{10}(\mathrm{VMR}) < -3.0$ in steps of $\Delta\log_{10}\mathrm{(VMR)} = 0.5$ for the three molecular species, at pressures of $p_1 = (1, 0.1, 0.01)$ bar and $p_2 = (10^{-3}, 10^{-4}, 10^{-5})$ bar, and at three base temperatures of $T_1 = (1000,1250,1500)$ K. These are representative of planet equilibrium temperatures in various cases of energy redistribution. $T_2$ was fixed to 500 K for non-inverted profiles, and to 1500 K for inverted profiles. All possible combinations of parameters were employed. The radiative transfer in the planet atmosphere was computed in the plane-parallel approximation and typical resulting models are shown in Figure~\ref{models}. As explained in Section~\ref{atm}, in order to set lower limits on the VMRs of molecules, we also assumed adiabatic lapse rates. In this case, $p_2$ is set by the altitude at which the atmosphere reaches $T_2 = 500$ K.

\begin{figure*}
\centering
\includegraphics[width=13cm]{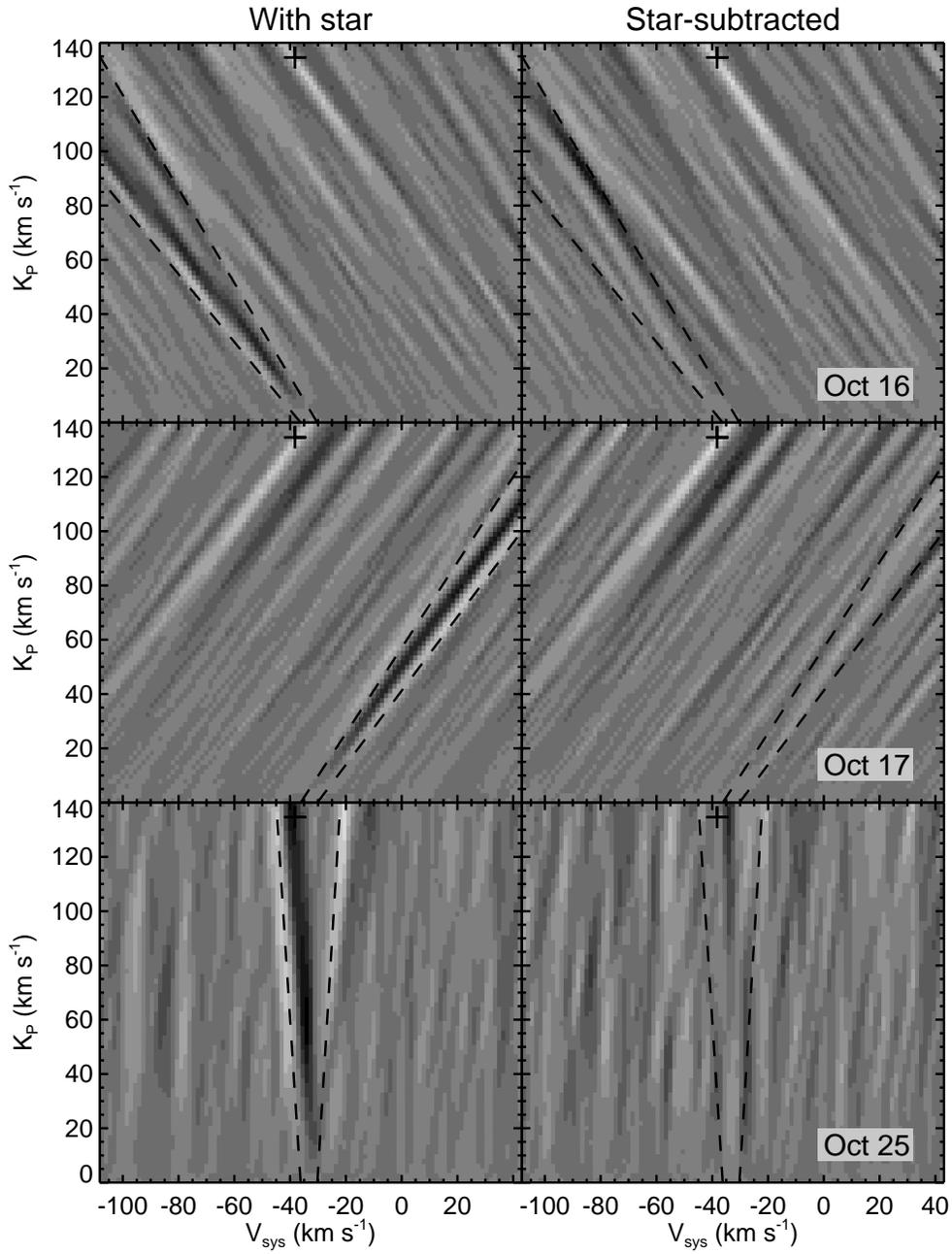}
\caption{{\it Left column}: Total cross-cross correlation signal for each night of observations, without removing the stellar absorption lines. {\it Right column}: Same as the left column, but after removing the stellar spectral features. In all panels a linear color scheme running from -7$\sigma$ (black) to +7$\sigma$ (white) is adopted. Crosses indicate the planet position measured by combining the nights of October 16 and 17. Black-dashed lines encompass regions of the parameter space significantly affected by stellar residuals. These result in extra cross-correlation noise potentially masking the planet signal in the third night, as discussed in Section~\ref{issue-data}.}\label{diag}
\end{figure*}

\subsection{Cross correlation}\label{ccorr}

In our observed wavelength range, the expected spectral signature of 51 Pegasi b consists of tens of carbon monoxide lines, plus a possible additional contribution from water vapor and/or methane, with a typical depth per-line of few parts in 10$^5$ with respect to the stellar continuum. Consequently, the individual lines will be in any case deeply buried in the noise, as each spectrum has a signal-to-noise ratio of typically 500. In order to detect the planet signature, the signals from all the lines are combined with a cross-correlation technique \citep{sne10,bro12}. Instead of directly cross-correlating the atmospheric models with the data, a template spectrum is constructed from the model by subtracting the baseline, clipping out the weakest lines and convolving the result to the CRIRES resolution. This is done because any broad-band signal has been removed by our data reduction technique. In addition, the line profiles in the model should match the instrumental profile in order to maximize the cross-correlation signal. Each spectrum is cross-correlated with the template using a lag-vector corresponding to radial velocities between --250 and +250 km s$^{-1}$, in steps of 1.5 km s$^{-1}$, and the result for each detector and each night is stored in a matrix with $x$- and $y$-dimension representing radial velocity and frame number (phase) respectively. Subsequently, the matrices of the four detectors are summed with equal weighting, resulting in a single cross-correlation matrix per night of observation. As an exception, detector 4 was discarded from the analysis during the third night (October 25, 2010), as explained in Section~\ref{issue-data}. For models with non-inverted (inverted) $T/p$ profiles, a positive cross-correlation signal indicates absorption (emission) in the planet atmosphere. At this stage the signal-to-noise is insufficient to detect the peak of the cross-correlation function per spectrum so we cannot derive the planet radial velocity curve on a frame-by-frame base. Instead, the cross-correlation signal needs to be combined over time by considering a range of maximum planet radial velocities ($K_\mathrm{P}$) corresponding to the possible system inclinations. In all cases we assume a circular orbit. For each value of $K_\mathrm{P}$, the planet radial velocity curve was computed and utilized to shift each cross-correlation function to the rest frame of the planet, via linear interpolation. Finally, the shifted cross-correlation series was summed over time, resulting in the total cross-correlation signal as function of systemic radial velocity ($V_\mathrm{sys}$) and maximum planet radial velocity. If 51 Pegasi b is detected, we should find a peak in the cross-correlation at the known systemic velocity of the system ($V_\mathrm{sys} = -33.25$ km s$^{-1}$), for the best matching inclination. Any spectral signature which is not accelerating with respect to the observer will appear at $K_\mathrm{P} = 0$, including residual telluric and stellar absorption features. More importantly, since the cross-correlation signal is summed in time by assuming a range of inclinations, strong telluric and stellar signals will also show residuals at $K_\mathrm{P} \ne 0$. These will be shifted in systemic velocity in a non-trivial way, depending on the phase range of the observations (see e.g.\ the left column of Figure~\ref{diag}). This is why telluric removal and stellar subtraction are crucial steps of our data analysis.

\section{The cross-correlation signal}\label{results}

The total cross-correlation signal, computed as explained in Section~\ref{ccorr}, is shown in Figure~\ref{diag} for each of the three nights of observations, using the planet model spectrum that best matches the data. This is selected from our large grid of models as explained in Section~\ref{atm} and corresponds to atmospheric absorption (non-inverted $T/p$ profile) of carbon monoxide and water, with VMRs of $10^{-4}$ and $3\times 10^{-4}$ respectively. On the nights of October 16 and 17, 2010, candidate signals from 51 Pegasi b are detected at the orbital velocities and systemic velocities listed in Table~\ref{tab-det}. Residuals from stellar lines appear as extra cross-correlation noise in a region of the parameter space approximately enclosed within the black dashed lines in Figure~\ref{diag}. The night of October 25 does not show any significant planet signal, and we discuss possible reasons for this in Section~\ref{discussion}.

\begin{figure}[th]
\epsscale{1.0}
\plotone{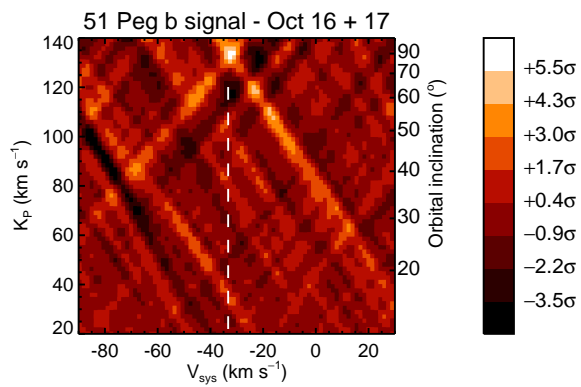}
\caption{Total cross-correlation signal obtained by combining the nights of October 16 and 17 and by applying a global phase shift of $\Delta\varphi = 0.0095$. The planet signal is recovered at the systemic velocity of 51 Pegasi ($V_\mathrm{sys} = -33.25$ km s$^{-1}$, white dashed line) and at a maximum planet radial velocity of $K_\mathrm{P} = (134.1\pm 1.8)$ km s$^{-1}$. Note that there is an uncertainty in linking the measured $K_\mathrm{P}$ (left vertical axis) to the orbital inclination $i$ (right vertical axis). This is due to the uncertainties in the stellar mass (see Section~\ref{mass-inc}).}\label{diag_tot}
\end{figure}

Figure~\ref{diag_tot} shows the cross-correlation signal of the nights of October 16 and 17 combined. A peak in the total cross-correlation signal is detected at the orbital radial velocity of $K_\mathrm{P}$ = (134.1 $\pm$ 1.8) km s$^{-1}$. Its significance of 5.9$\sigma$ is computed by dividing the peak cross-correlation value by the standard deviation of the noise. This points to a high orbital inclination, as discussed in Section~\ref{mass-inc}. We discuss how this signal compares to that obtained using models with pure water and pure carbon monoxide in Section~\ref{atm}. A phase shift of $\Delta\varphi = 0.0095$ needs to be applied to the orbital parameters of the system given by \citet{but06} in order to match the signal to the known systemic velocity of 51 Pegasi ($K_\mathrm{sys} = -33.25$ km s$^{-1}$). This shift is well within the $1\sigma$ uncertainty range of $\Delta\varphi = \pm 0.012$ derived from the same orbital solution.

\begin{figure*}[ht]
\centering
\includegraphics[width=15cm]{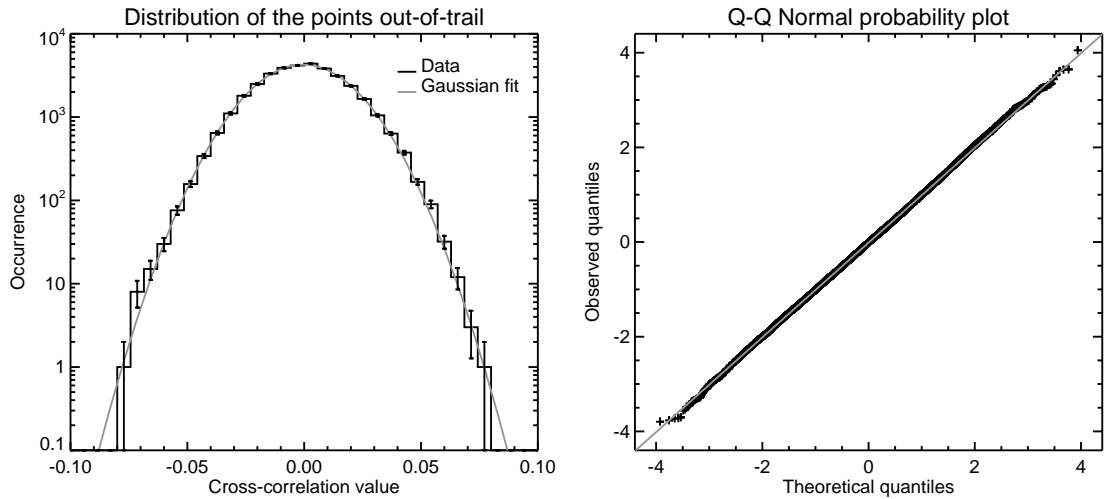}
\caption{{\it Left panel:} the distribution of the cross-correlation values for the nights of October 16 and 17, 2010 is shown in logarithmic scale, with 1$\sigma$ error bars corresponding to the square root of the bin occurrences. Overplotted in grey is the expected distribution for Gaussian noise. {\it Right panel:} the quantiles of a Normal distribution are plotted against the quantiles of our measured cross-correlation values (Q-Q plot). In both panels, the observed and theoretical distributions agree across the entire range which can be probed ($\pm4\sigma$), which is set by the number of cross-correlation points available.}\label{cc-dist}
\end{figure*}

Since one third of the data do not show the planet signal, it is particularly important to robustly assess the statistical significance of the signal in the first two thirds of the data. For this purpose, the properties of the cross-correlation noise were studied by comparing the distribution of the cross-correlation values with that of a Gaussian distribution. This is shown by the left panel of Fig.~\ref{cc-dist}. In addition, in the right panel the quantiles of the observed distribution are plotted against the quantiles of a Normal distribution \citep[Q-Q plot,][]{wilk68}. In both cases, the two distributions agree within $\sim$4$\sigma$, a limit set by the finite number of points available. It is not surprising that the cross correlation has Gaussian noise properties, as non-Gaussian noise originating from a variable signal-to-noise is avoided by normalizing the data by the square of their standard deviation (see Section~\ref{telluric}). Moreover, systematic noise is also strongly scaled down by the fact that the template spectrum contains tens of well-spaced molecular lines, and therefore the cross-correlation spans large portions of the data.

\begin{figure}[ht]
\epsscale{1.0}
\plotone{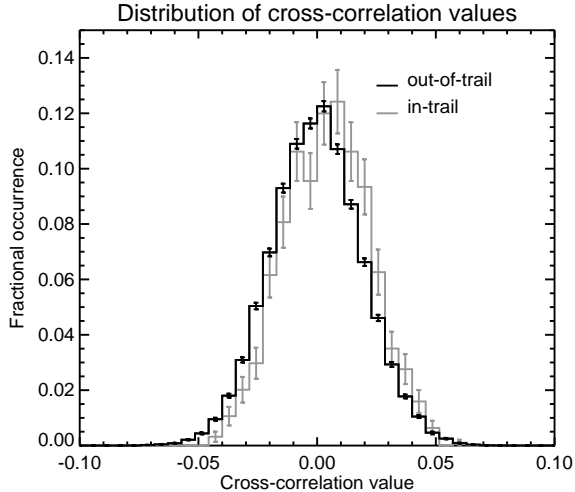}
\caption{Normalized distribution of the cross-correlation values out of the planet trail (black line) compared to the same distribution for the in-trail cross-correlation values. The in-trail distribution is centered to higher values than the out-of-trail distribution at a 5.4$\sigma$ confidence level (see Figure~\ref{welch_t}). The error bars corresponds to the square-root of the bin occurrences.}\label{in-out}
\end{figure}

\begin{figure}[ht]
\epsscale{1.0}
\plotone{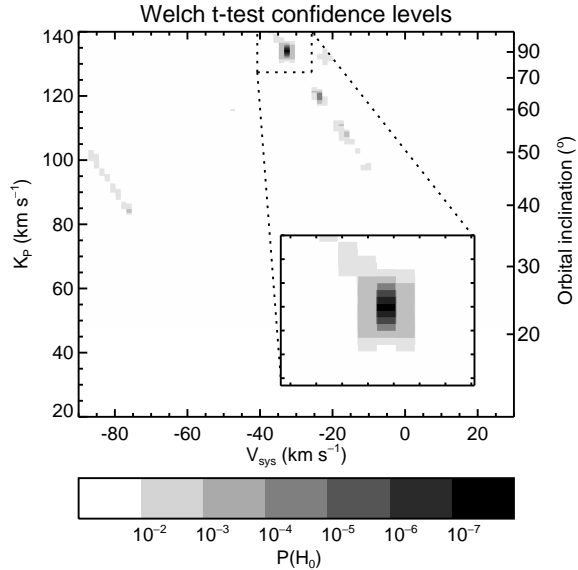}
\caption{Results of a generalized t-test on the null hypothesis that the in- and out-of-trail cross-correlation values are drawn from the same parent distribution. This test is performed for the same combination of systemic velocities and planet orbital velocities as in Figure~\ref{diag_tot}. It indicates that the null hypothesis $H_0$ can be rejected at the 5.4$\sigma$ confidence level for the planet signal. Note that there is an uncertainty in linking the measured $K_\mathrm{P}$ (left vertical axis) to the orbital inclination $i$ (right vertical axis). This is due to the uncertainties in the stellar mass (see Section~\ref{mass-inc}).}\label{welch_t}
\end{figure}

In Figure~\ref{in-out} we compare the distribution of the cross-correlation values inside the planet trail (i.e.\ those around the planet radial velocity for the derived $K_\mathrm{P}$) with that of the points outside the planet trail. The in-trail distribution is shifted towards higher values than the out-of-trail distribution. In order to quantify this, a Welch t-test on the two samples was performed \citep{wel47}, with the null hypothesis that they are drawn from the same parent distribution. Based on the results of the test, the null hypothesis is rejected at the 5.4$\sigma$ confidence level, in line with our initial estimate based on pure signal-to-noise calculation. Moreover, Fig.~\ref{welch_t} shows the Welch t-test performed for the full range of $V_\mathrm{sys}$ and $K_\mathrm{P}$ of Fig.~\ref{diag_tot}. Remarkably, no other region of the parameter space shows significant cross-correlation signal, except for two slanted lines converging towards the planet signal. This is expected from a real planet signal, because the method used for summing the cross-correlation functions in time naturally produces signals fading away from the planet position. For this reason two or more observations taken at different phases are much more effective in constraining the planet parameters than a single observation.

\section{The non-detection on the third night}\label{discussion}
As explained in Section~\ref{results}, the third night of observations does not show any sign of an absorption signal from the dayside of 51 Pegasi b. This is a concern because of the statistical robustness of the detection during the first two nights (see Section~\ref{results}), which makes it unlikely that time-correlated noise could mimic the planet signal. Therefore, it is important to investigate the possible causes of this discrepancy. We discuss below i) possible issues with the instrument, data and/or data analysis for the third night, and ii) possible astrophysical causes. 

\subsection{Issues with instrument, data and data analysis}\label{issue-data}
The night of October 25 experienced a relatively good seeing and a longer coherence time than the other two nights of observations, resulting in better performances of the MACAO system and therefore in a higher signal to noise compared to the first two nights (by 5-15\% per spectrum). However, the stability of the CRIRES spectrograph was significantly worse, causing the spectra to drift by $\sim$1.5 pixels in the dispersion direction, whereas during the other two nights these shifts were only 0.2--0.5 pixels. The cause of this drift is likely associated with thermal instabilities of the cryogenic system. The temperature of the pre-optics system, grating stabilizer and grating itself changed by about 1.5 K during the 5 hours of observation on October 25, whereas during the other two nights it was stable at the $< 0.05$ K level. This drift in temperature is indeed correlated with the drift in spectral position on the detector. Since the spectra need to be aligned to a common reference frame (see Section~\ref{datared}), the wavelength drift has a negative impact on the quality of the data. We were forced to discard detector 4 from the analysis of the third night, because it suffers from gain variations between neighbouring columns (the so-called odd-even effect). While with sub-pixel shifts this is not a significant problem (spectral features roughly fall on the same pixel), for shifts of one pixel or more the odd-even effect introduces strong time-correlated noise in the data of up to several percent. In addition, general flat-field inaccuracies are affecting the data more significantly. 

Despite the instrumental issues during the third night, the expected significance of the detection in the third night is $\sim3.8\sigma$, based on the measured signal-to-noise, the total number of spectra, and the fraction of data lost because of the exclusion of detector 4. This means that, taking into account statistical fluctuations, there is a probability of 20\%, 10\% and 5\% to measure a signal less than 2.5$\sigma$, 2.2$\sigma$ and 1.8$\sigma$ respectively during the third night.

Another potential issue concerns residual stellar spectral features. As explained in Section~\ref{ccorr} and shown in Figure~\ref{diag}, even after subtracting a model stellar spectrum from the data, residual cross-correlation noise is seen. This is likely due to changes in the spectral response profile during the night and to inaccuracies in the alignment. This is particularly an issue during the third night, because the system was observed at phase $\varphi\sim 0.5$, when the planet spectral lines overlap with those of the star (i.e.\ planet and star have almost the same radial velocity). In contrast, during the nights of October 16 and 17 the planet and the star have a different radial velocity, causing this residual cross-correlation noise to be located in regions of the parameter space far away from the planet position, and therefore not significantly influencing the measured signal. The effects of the stellar residuals are shown in Figure~\ref{diag}, before and after the subtraction of the stellar spectrum (left and right columns respectively). During the third night of observations the expected planet signal falls clearly on top of the residual stellar cross-correlation (bottom-left panel). After the subtraction of the stellar spectrum residual signal is still visible (bottom-right panel). It is possible to filter out these stellar residuals by masking 5--7 pixels in the cross-correlation matrix, around the stellar radial velocity. In this way another $\sim$~40\% of the planet signal is removed, as estimated by injecting a model spectrum in the data at 3$\times$ the nominal level and at the previously determined $K_\mathrm{P}$ and re-running the data analysis. Even after masking, no significant planet signal is detected. Although stellar residuals cannot explain the complete absence of a planet signal, they certainly damp part of it. This is another important factor that differentiates the third night from the other two.

\subsection{Astrophysical causes}\label{issue-astro}

We also explored which possible astrophysical effects may cause the non-detection during the third night. First of all, the planet is observed around phase $\varphi = 0.5$, meaning that if it was a transiting system the planet would be eclipsed by the star, potentially explaining why no signal was detected. However, photometric monitoring of the 51 Peg system shows that the planet does not transit \citep{hen97,wal06}. For a system to exhibit secondary eclipses without transits, we calculated that the orbital eccentricity needs to be at least $e = 0.13$ for a stellar radius of $R_\mathrm{S} = (1.237 \pm 0.047) R_\odot$ \citep{bel09} and a planet radius of 1 $R_\mathrm{Jup}$. This is clearly excluded by the radial velocity data of the star \citep{but06}. 

Small deviations from the best-fit orbital solution to these radial velocity data can influence the radial velocity and timing of the expected planet signal, which could cause the non-detection during the third night. All the orbital solutions available in the literature agree over the fact that the radial velocity data of 51 Pegasi are equally well fit by a circular orbit as by a slightly eccentric orbit \citep[e.g.][]{may95,but06}. \citet{but06} obtained a best solution with $e = (0.013\pm0.012)$ and $\omega \approx 58^\circ$. We tested various eccentric solutions in order to retrieve the total cross-correlation signal and at the same time the $\chi^2$ of the orbital solution to the stellar radial velocity data \citep{but06}. Assuming an eccentricity of $e = 0.025$ and $\omega \sim 180^\circ$, which is more than 2$\sigma$ away from the best-fit solution of \citet{but06}, it was only possible to shift the signals of the third night by $\sim$~4  km s$^{-1}$, whereas the closest cross-correlation peak in the third night can be aligned to the other two only if shifted by about 10 km s$^{-1}$. This means it is not possible to explain the non-detection by assuming an eccentric solution.

When discussing astrophysical causes, there is also the possibility that the signal from the atmosphere itself is changing, either due to an orientation effect or due to to temporal changes. At orbital phase $\varphi = 0.5$, the orientation of the planet is almost precisely on the sub-stellar point, meaning that a relatively hotter atmosphere is observed. Although the presence of an inversion layer in hot-Jupiter atmospheres is still not well understood, stellar irradiation is thought to play a role \citep{for08}. However stellar activity \citep{knu10} and different elemental abundances \citep{mad12} have also been suggested as possible causes. Higher temperatures mean that some molecular species remain in the gas phase, and are capable of absorbing radiation high up in the atmosphere, causing a thermal inversion and spectral features in emission. If the atmosphere of 51 Pegasi b would be partially inverted around the sub-stellar point, this may cancel out part of the signal. In the limit of a nearly isothermal dayside atmosphere, as it seems to be the case for some hot Jupiters \citep[see e.g.][]{cro12}, no signal would be detected. However, the illumination of the visible planet disk changes only by $\sim$20\% during these observations, requiring a large temperature gradient as function of longitude and a significant contribution in absorption from the night-side, which is in contrast with an inefficient heat transport required by the presence of a much hotter sub-stellar spot. Moreover, such a large gradient would also show up in our data as a phase-dependent cross-correlation signal, something that we do not detect. This leaves only the possibility of temporal variations, the details and timescale of which are largely unknown even for hot Jupiters. Although theoretical studies predict weather to be important in hot Jupiters and brown-dwarfs \citep{sho12}, with only three nights of observations we are reluctant to attribute the absence of the planet signal to exoplanetary weather.

\section{Characterization of exoplanet 51 Peg b}\label{parameters}

In the previous sections we showed that our observations resulted in an apparent planet signal in the first two nights, but no detection in the third night. We have discussed several potential causes of the non-detection, including genuine astrophysical causes, none of which was robustly proven to be associated to the lack of signal. Even though we cannot confidentially exclude the possibility that the signal observed in the first two nights is due to a yet unknown and unfortunate instrumental effect, in this Section we assume that the signal is genuine and proceed with the characterization of 51 Pegasi b.

\subsection{Planet mass and system inclination}\label{mass-inc}

The known radial velocity amplitude of the host star, $K_\mathrm{S}$~=~(55.94~$\pm$~0.69) m s$^{-1}$ \citep{but06}, combined with our measured planet radial velocity amplitude of $K_\mathrm{P} = (134.1\pm 1.8)$ km s$^{-1}$, gives a star/planet mass ratio of $M_\mathrm{S}/M_\mathrm{P} \equiv K_\mathrm{P}/K_\mathrm{S} = (2397 \pm 44)$. With a stellar mass of $M_\mathrm{S}$ = (1.05 $\pm$ 0.04) $M_\odot$, as determined by \citet{tak07}, it translates into a planet mass of $M_\mathrm{P}$ = (0.46 $\pm$ 0.02) $M_\mathrm{Jup}$. The orbital velocity of the planet can be computed through Kepler's Third Law, and it is equal to $V_\mathrm{P}$ = (133.7 $\pm$ 1.8) km s$^{-1}$. The sine of the inclination, $\sin i = K_\mathrm{P}/V_\mathrm{P} = (1.003 \pm 0.019)$, implies 1$\sigma$, 2$\sigma$ and 3$\sigma$ lower limits on the orbital inclination of 79.6$^\circ$, 74.7$^\circ$ and 71.0$^\circ$ respectively. Observations conducted with the MOST satellite show that the planet is not transiting \citep{wal06}. Using the planet and stellar radii given in Sect.~\ref{issue-astro}, this sets an upper limit on the system inclination of $i_\mathrm{max} = (82.2 \pm 0.3)^\circ$.

\subsection{The atmosphere of 51 Pegasi b}\label{atm}

The first information which is derived from the cross-correlation signal is that the atmosphere of 51 Pegasi b is not thermally inverted in the pressure range probed by these observations. This is because significant planet signal is only observed when using models with non-inverted $T/p$ profiles (atmospheric absorption). In addition, significant cross-correlation signals are obtained not only from models with pure carbon monoxide, but also from models with pure water, at a $\le 5.7\sigma$ and $\le 4.1\sigma$ level respectively. Details of these single-molecule detections are given in Table~\ref{tab-single-det}. No significant signal from a pure methane model is observed. This suggests that, while in our previous study of $\tau$ Bo\"otis b the cross-correlation signal was entirely due to CO, here there is a significant contribution from H$_2$O lines. In order to take this into account, another set of atmospheric models was computed by including both molecules, for the same range of pressure and temperature as in Section~\ref{atmmodel}. Further information about the structure of the planet atmosphere can be derived by comparing the cross-correlation signal retrieved by using this additional grid of atmospheric templates. Note that, since the data analysis removes broad-band spectral features such as the slope of the planet continuum and the wings of broad lines, the cross correlation is only sensitive to narrow spectral features, such as the flux ratio between the planet continuum and the core of the molecular lines. The comparison between models and data can be done in the following two ways. Firstly, different model spectra, exhibiting varying line ratios, can be cross-correlated with the data. The model spectrum that matches the data the best will show the the strongest signal. Secondly, the amplitude of the model spectrum can be compared with the dayside spectrum of 51 Pegasi b by subtracting the model spectrum from the data early on in the analysis chain, before telluric removal. This comparison must account for the dilution by the starlight, which results in the following normalized planet spectrum:
\begin{equation}
F_\mathrm{norm}(\lambda) = \frac{F_\mathrm{model}(\lambda)}{F_\mathrm{star}(\lambda)} \left(\frac{R_\mathrm{P}}{R_\mathrm{S}}\right)^2,
\end{equation}
where $F_\mathrm{model}$ is the model spectrum, $F_\mathrm{star}$ the stellar spectrum (for which we use the NextGen low-resolution model for a 51 Peg-like star with $T_\mathrm{eff}$ = 5800 K, $\log g$ = 4.5, and solar metallicity) and $R_\mathrm{P}$ and $R_\mathrm{S}$ the planet and stellar radii respectively. This scaled model is reversed and injected in the data at the measured planet position, before telluric removal. As a result, the cross-correlation signal will be cancelled out when the model matches the data. However, the fact that the planet is not transiting the host star adds significant uncertainty to this second test, because $R_\mathrm{P}$ is unknown. Where not specified, 51 Pegasi b is assumed to have the size of Jupiter.

It is important to realize a degeneracy exists between molecular abundances and atmospheric lapse rate. The higher the VMR, the higher the altitude at which a molecule absorbs. This generally translates to a larger temperature difference between the layer in which the planet continuum is formed and the altitude at which a molecule absorbs in the core of a line. However, this difference is also related to the atmospheric lapse rate, i.e.\ the rate at which the temperature changes with pressure. Since the temperature difference determines the contrast between the continuum and the core of the lines, the observed strength of the high-resolution planet signal is affected by this degeneracy, which cannot be solved with a single observation in a limited wavelength range. In this work, the lapse rate was fixed to the adiabatic value (i.e.\ the maximum allowed in a planet atmosphere, corresponding to the onset of convection), and the base temperature $T_1$ was chosen to be roughly the planet equilibrium temperature (see Sect.~\ref{atmmodel} for details). This results in a maximum planet signal for a given molecular abundance and therefore sets lower limits in the molecular VMRs.

Among the large set of models generated as explained in Sect.~\ref{atmmodel}, we selected the best-matching dayside model spectrum of 51 Pegasi b by requiring, at the same time, the highest signal-to-noise of the total cross-correlation signal and zero residual cross-correlation signal after subtracting the model spectrum from the data. A range of $\pm 1\sigma$ around these two values is explored in order to allow noise fluctuations. This criterium resulted in a model with $T_1 = 1250$ K, $p_2 = 10^{-4}$, VMR(CO) = 10$^{-4}$, VMR(H$_2$O) = 3$\times$10$^{-4}$, meaning that roughly 40\% of the signal is due to H$_2$O molecular lines. This is not surprising, given the relatively lower equilibrium temperature of 51 Pegasi b as compared to $\tau$ Bo\"otis. Below 1500 K carbon monoxide starts to be significantly less abundant than water for solar C/O ratios \citep{mad12}. Our detection of water together with carbon monoxide, assuming the signal is genuine, is therefore in line with a relatively cooler hot Jupiter. Note however that the relative abundance of water with respect to carbon monoxide is only weakly constrained by these observations, with the 1$\sigma$ range being 0.05~$<$~VMR(H$_2$O)/VMR(CO)~$<$~140.

Note that, intuitively, atmospheric models containing both water and carbon monoxide would produce a cross-correlation signal which is roughly the quadrature sum of that obtained from models with pure H$_2$O or pure CO, combining in this case to $\sim 7\sigma$ of significance. However, in practice, this does not happen, because the two cross-correlation signals cannot be considered to be independent. Superposition of CO and water lines (see e.g. Figure~\ref{models}, right panel), and a mismatch between the real planet spectrum and the model templates used, can cause the total signal to deviate from a simple quadrature sum of that of the individual gases. Also, the combined model redistributes the noise content of the data. This means that the combined signal may deviate from the value expected from the individual gases with a probability distribution having a standard deviation of $1\sigma$.

\section{Conclusions and future prospects}\label{concl}
By observing the orbital motion of 51 Pegasi b directly using high-resolution spectra, we have likely been able to disentangle the spectral fingerprints of the planet's atmosphere from telluric and stellar contamination, and reveal significant absorption from carbon monoxide and water vapour in its dayside hemisphere using the first two nights of observations. This was achieved by studying the cross-correlation of the data with a model planet atmosphere. In this way the radial velocity of the planet is measured directly, allowing us to determine the planet/star mass ratio and the system inclination. This analysis suggests that 51 Pegasi b has a mass of $M_\mathrm{P}$ = (0.46 $\pm$ 0.02) $M_\mathrm{Jup}$ and an orbital inclination between 79.6$^\circ$ and 82.2$^\circ$, with the upper limit set by the non-detection of transits in photometric monitoring of the host star. By comparing the cross-correlation signal obtained with a range of atmospheric models, we estimate minimum molecular abundances of VMR(CO) = 10$^{-4}$ and VMR(H$_2$O) = 3$\times$10$^{-4}$.

However, the absence of the signal in the third night of observations is puzzling. Based on a conservative signal-to-noise estimate, there is a 20\% probability of observing a signal less than 2.5$\sigma$ in the third night purely due to noise fluctuations. Moreover, the stability of the CRIRES spectrograph was much worse during the third night of observations, so that possible flat-field inaccuracies affect a larger portion of the data. Furthermore, residual cross-correlation signals from stellar features are present at the expected planet position in the night of October 25. We also investigated possible astrophysical causes of the lack of signal, i.e.\ the presence of a mild temperature inversion at the sub-stellar point or atmospheric variability due to weather. However, none of the possible instrumental and astrophysical causes can be confidentially linked to the non-detection, meaning that future observations of 51 Pegasi will be essential in confirming the signal presented in this work.

Ground-based high-resolution spectroscopy in the near-infrared is a powerful tool for investigating hot Jupiter atmospheres. While the interpretation of low-resolution studies during transits or secondary eclipses is often controversial, due to the fact that molecular features superimpose across the same wavelength range, this technique unambiguously identifies molecular signatures thanks to the use of a cross-correlation technique with template models. Moreover, it delivers the planet radial velocities directly and does not require -- when applied to dayside spectroscopy -- a planet to transit. The only requirement is that the planet must move significantly in radial velocity (by 10 -- 20 km s$^{-1}$) during the observations. 

The main limitations of this technique are currently the degeneracy between atmospheric structure and molecular abundances, and the need for very bright targets. The former can in principle be solved by observing at multiple wavelengths and by detecting the signatures of different molecular species, a goal that is within the reach of our current methods and of the available instrumentation. The latter will certainly benefit from the next-generation of telescopes with mirrors of 30 m or more, and possibly from spectrographs capable of observing larger wavelength ranges at once, as the total cross-correlation signal is approximately proportional to the square root of the number of strong molecular lines observed.  

Future developments of high-resolution spectroscopy may involve the study of the shape of the cross-correlation peak during transit, capable of revealing atmospheric rotation and global circulation patterns. Furthermore, by accumulating enough signal-to-noise, it will be possible to study the phase curve of these hot Jupiter and trace changes in the molecular abundances and in the atmospheric chemical processes between their morning and evening sides. Eventually, by employing a 39-m class telescope such as the E-ELT, it may be possible to detect oxygen in the atmosphere of an Earth-like planet orbiting in the habitable zone of an M-dwarf star \citep{sne12}. This would be a considerable jump forward towards the detection of biosignatures in Earth analogs.

\acknowledgments

We are thankful to the ESO staff of Paranal Observatory for their support during the observations. This work is based on data collected at the European Southern Observatory (186.C-0289). I. S. acknowledges support by VICI grant no. 639.043.107 from the Netherlands Organisation for Scientific Research (NWO). S.A. acknowledges support by a Rubicon fellowship from the NWO, and by NSF grant no. 1108595.

{\it Facilities:} \facility{VLT:Antu (CRIRES)}.

\newpage

\begin{table*}
\begin{center}
\begin{tabular}{|c|cccc|}
\tableline
Night & $K_\mathrm{P}$ & $V_\mathrm{sys}$ & Phase & Significance \\
      & (km s$^{-1}$) & (km s$^{-1}$) & range  &\\
\tableline
2010-10-16 & 122.3$^{+15.9}_{-24.8}$ & -25.7$^{+15.0}_{-10.3}$ & 0.36--0.42 & 4.5$\sigma$ \\[3pt]
2010-10-17 & 134.7$^{+13.7}_{-21.3}$ & -33.2$^{+9.8}_{-10.3}$ & 0.60--0.66 & 4.7$\sigma$ \\[3pt]
2010-10-25 & \nodata & \nodata & 0.49--0.54 & $<$2.5$\sigma$ \\[3pt]
16 + 17    & 134.1$\pm$1.8 & -33.2$\pm$1.5 & n.a. & 5.9$\sigma$ \\[7pt]
\tableline
\end{tabular}
\caption{Details of the single-night detections of 51 Pegasi b. The planet orbital velocity, the systemic velocity, the observed range in planet orbital phase and the significance of the detection are given. The last row shows the nights of October 16 and 17 combined. A global phase shift of $\Delta\varphi$ = 0.0095 is applied to the data in order to make the systemic velocity of the night of October 16 and 17 combined to coincide with the systemic velocity of 51 Pegasi.}\label{tab-det}
\end{center}
\end{table*}

\begin{table*}
\begin{center}
\begin{tabular}{|c|ccc|}
\tableline
Molecule & $K_\mathrm{P}$ & $V_\mathrm{sys}$ & Significance \\
      & (km s$^{-1}$) & (km s$^{-1}$) &  \\
\tableline
H$_2$O & 136.9$\pm$2.8 & -34.9$^{+1.4}_{-2.0}$ & 4.1$\sigma$ \\[3pt]
CO     & 134.2$\pm$1.8 & -33.2$\pm$1.5 & 5.7$\sigma$ \\[7pt]
\tableline
\end{tabular}
\caption{Details of the cross-correlation signals obtained by employing models with pure water or pure carbon monoxide, for the nights of October 16 and 17 combined. The planet orbital velocity, the systemic velocity and the significance of the detection are given.}\label{tab-single-det}
\end{center}
\end{table*}

\end{document}